\documentclass[aps,pre,longbibliography,twocolumn]{revtex4-2}

\usepackage{graphicx}

\usepackage{dcolumn}

\usepackage{amsmath}
\usepackage{bm}
\usepackage[normalem]{ulem}
\usepackage{color}

\usepackage{epstopdf}

\newcommand{\tri}{\triangle}

\newcommand{\beq}{\begin{equation}}
\newcommand{\eeq}{\end{equation}}
\newcommand{\beqn}{\begin{eqnarray}}
\newcommand{\eeqn}{\end{eqnarray}}
\newcommand{\pp}{\partial}
\newcommand{\dd}{{\rm d}}

\newcommand{\fig}{Fig.\ }

\newcommand{\la}{\langle}

\newcommand{\ra}{\rangle}

\begin{document}

\title{Scaling Law and Universal Drop Size Distribution of Coarsening in  Conversion-Limited Phase Separation}

\author{Chiu Fan Lee}
\email[Electronic address: ]{c.lee@imperial.ac.uk}
\affiliation{Department of Bioengineering, Imperial College London, South Kensington Campus, London SW7 2AZ, United Kingdom}

\date{\today}


\begin{abstract}
Phase separation is not only ubiquitous in diverse physical systems,  but also plays an  important organizational role  inside biological cells. However, experimental studies of intracellular condensates (drops with condensed concentrations of  specific collections of proteins and nucleic acids) have challenged the standard coarsening theories of phase separation.
Specifically, the coarsening rates  observed are unexpectedly slow for many intracellular condensates. Recently, Folkmann, et al. [Science {\bf 373}, 1218 (2021)] argued that
the slow coarsening rate can be caused by the slow conversion of a condensate constituent between the state in the dilute phase and the condensate state. A consequence of this conversion-limited picture is that standard theories of  coarsening in phase separation (Lifshitz-Slyozov--Wagner Ostwald ripening and drop coalescence  schemes) no longer apply. Surprisingly, I show here that  
 the model equations of conversion-limited phase separation can instead be mapped onto a  grain growth model in a single-phase material in three dimensions. I further elucidate the universal coarsening behavior in the late stage  using analytical and  numerical methods.

\end{abstract}

\maketitle

\section{Introduction}
Phase separation 
is a ubiquitous phenomenon in nature: 
from the separation of  quark matter with distinct baryon densities in the  early universe \cite{schwarz_ann03,randrup_prc09}, to the
everyday occurrence of dew and fog  on earth today. 
Besides being integral to our understanding of diverse physical systems, phase separation also plays an important organizational role in living systems: many  protein-nucleic acid condensates (i.e., drops with condensed concentrations of  specific collections of proteins and nucleic acids) exist intracellularly \cite{hyman_annrev14,shin_science17}. This recent revelation led to an intense interest of cellular phase separation from cell biologists and biophysicists \cite{brangwynne_natphys15,banani_natrev17,berry_rpp18, weber_rpp19}. One of the key outstanding questions in this emergent field is: Why do the coarsening rates of many condensates observed in cells seem negligible?

 Coarsening of condensates refers to the evolution from an emulsion of  polydisperse condensates to a single condensate co-existing with the dilute phase, as dictated by thermodynamics. 
Coarsening in the late-stage can take on two forms: (i) the transfer of material from small drops to big drops through the dilute phase in a process known as Ostwald ripening, 
and (ii) the coalescence of diffusing drops when they encounter each other \cite{weber_rpp19}. To account for the slow condensate coarsening rates observed in experiments, various proposals have been advocated: driven  chemical reactions that convert constituent proteins between
 a soluble form and a phase separating form can counter Ostwald ripening \cite{zwicker_pnas14,zwicker_pre15,wurtz_prl18},
 and intracellular visco-elastic networks (e.g., the cytoskeletal networks) can limit the growth of condensates through mechanical suppression \cite{feric_natcellbiol13,style_prx18,rosowski_natphys20,
 zhang_prl21}, and render the condensates' dynamics sub-diffusive, which  slows down coalescence \cite{lee_natphys21}.

More recently, 
drawing inspiration from the rugged energy landscape picture that explains the slow elongation rates observed in  amyloid fibrillization \cite{lee_pre_elongation09, straub_annrev11,schmit_jcp13},
a conversion-limited scheme was proposed to model the coarsening dynamics of P granules -- a type of cellular condensates found in the germ cells of the nematode {\it Caenorhabditis elegans}   \cite{folkmann_bio21}. Indeed, in many in vitro amyloid fibrillization experiments with no driven  chemical reactions, no visco-elastic networks limiting fibrillar growth, and no sub-diffusive behavior, unexpectedly slow elongation rates are nonetheless observed
\cite{lomakin_pnas97,scheibel_pnas04,ban_jmb04,collins_plosbiol04,
knowles_pnas07}. 
To rationalize these findings, a rugged energy landscape picture was proposed in \cite{lee_pre_elongation09}, in which the  many local free energy minima 
 arise from the various suboptimal conformations that a fibrillizing protein can be stuck in before achieving the minimal free energy state, which corresponds to the fully integrated fibrillar form. The slow fibril elongation rates observed can thus be explained by the generically slow   ``diffusion" over a rugged energy landscape  \cite{zwanzig_pnas88}.
 
\begin{figure}
	\begin{center}
		\includegraphics[scale=.72]{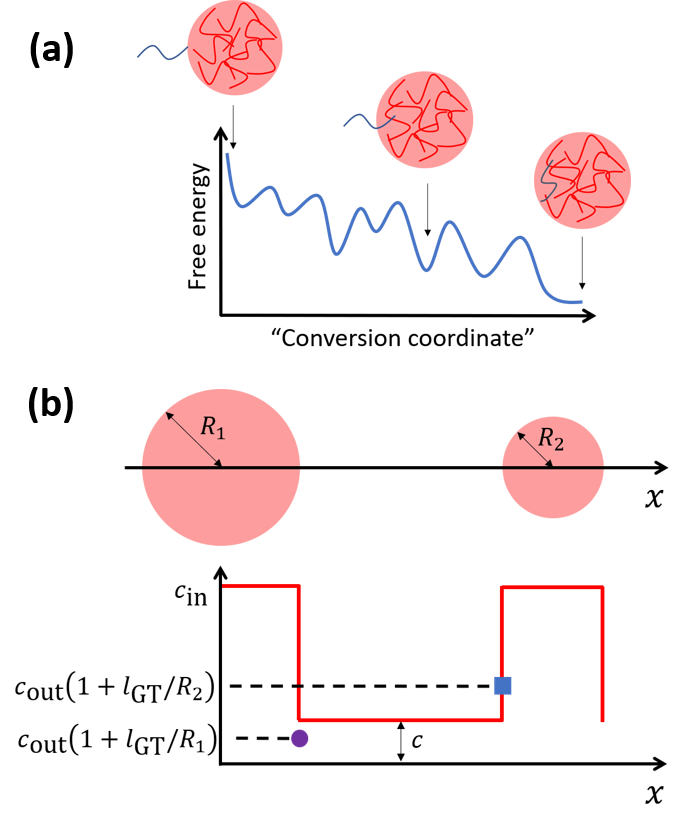}
	\end{center}
	\caption{Coarsening in the conversion-limited scheme. (a) A schematic of the rugged energy landscape that a monomer (blue) has to traverse through to become fully incorporated into the condensed drop (pink). 
	Here, the vertical axis depicts the free energy of the blue polymer-red drop system constrained to the partially converted state along the abstracted conversion coordinate.
	For a polymer with multiple binding patches, the rugged-energy landscape can arise from the sequential binding-unbinding events, many of them frustrated, that the polymer has to go through before being integrated into the drop.
Since diffusion over a rugged-energy landscape can be very slow \cite{zwanzig_pnas88}, the conversion step can become rate-limiting in coarsening. If this is the case, the concentration in the dilute phase, $c$, can be assumed to be constant throughout the dilute phase due to the fast molecular diffusion. (b) In a two-drop system (top figure), the concentration along the $x$ axis (red lines) according to the conversion-limited scheme is shown in the bottom figure. Due to the surface tension-induced Gibbs-Thomson relation \cite{weber_rpp19}, the equilibrium concentration outside the big drop (purple circle) is lower than that of the small drop (blue square). The mismatched concentrations outside the two drops lead an outflux of material from the small drop and an influx into the big drop, resulting in the coarsening of the system. Figures adapted from \cite{folkmann_bio21}.
	}
	\label{fig:cartoon}
\end{figure}

  In cellular condensates, the constituents (proteins and nucleic acids) can be thought of as polymers with multiple, potentially unspecific, binding sites. Therefore, it was argued in \cite{folkmann_bio21} that   a condensate constituent in the dilute phase also has to go through a series of local minima (e.g., a series of suboptimal binding configurations) before being fully incorporated into the condensate (\fig \ref{fig:cartoon}a). As a result, a corresponding rugged energy landscape picture can potentially 
  explain the slow coarsening rates observed.

An immediate consequence of
this conversion-limited picture is that the standard theories of coarsening
are no longer  valid. 
In this paper, I will first review in detail the derivation of the model equations that describe  coarsening in the conversion-limited scheme as proposed in \cite{folkmann_bio21}.  
I will then show that the model equations can be mapped onto a grain growth model in a single-phase material in three dimensions \cite{hillert_actamet65}. 
I further elucidate the universal coarsening behavior in the late stage  using analytical and  numerical methods.

\section{Model equations}
To proceed analytically, I will use a simplified single-component model, in which a single  concentration captures effectively the aggregate concentrations of the condensate constituents. 
In an emulsion of polydisperse drops with a uniform inner concentration $c_{\rm in}$, the equilibrium concentration outside a drop of radius $R$ is $c_{\rm out}(1+l_{_{\rm GT}}/R)$, where $c_{\rm out}$ is the concentration outside a flat interface ($R \rightarrow \infty$), and $l_{_{\rm GT}}$ is a length scale that accounts for the Gibbs-Thomson relation arising from the surface tension-induced Laplace pressure acting on the drops (\fig \ref{fig:cartoon}b) \cite{weber_rpp19}. 
In the conversion-limited regime, the slow process is the conversion of a constituent molecule into and out of a drop at the interface, while molecular diffusion in the dilute phase is the fast process that renders the concentration in the dilute phase, $c$, constant throughout. By the principle of mass conservation, $c$ is given by
\beq
\label{eq:c}
c(t) = \frac{c_{\rm tot}V -c_{\rm in}V_{\rm drops}(t)}{V-V_{\rm drops}(t)}
\ ,
\eeq
where $c_{\rm tot}$ is the total solute concentration, $V$ is the volume of the system, and $V_{\rm drops}(t)=\frac{4 \pi}{3} \sum_{i=1}^N R_i(t)^3$ is the total drop volume in the system with $N$ being the number of drops. 

To achieve the equilibrium state of having a single drop co-existing with the dilute phase, the system inevitably coarsens. To consider the universal behavior in the late stage, I will focus exclusively on coarsening by Ostwald ripening under the conversion-limited scheme, and justify the neglect of coarsening by drop coalescence later. 

The thermodynamic drive towards equilibrium is caused by the mismatch between the radius-dependent equilibrium concentrations  outside the drops given by
\beq
c_{\rm out}\left(1+\frac{l_{_{\rm GT}}}{R} \right)\ ,
\eeq 
and the  concentration of the dilute phase, $c(t)$ (\fig \ref{fig:cartoon}b). Here, I assume that the mismatch is small  so that the resulting material flux is proportional to the concentration difference. 
Specifically,  the flux, $J_R(t)$, into (when positive) and out of (when negative) a drop of radius $R$ is given by
\beq
J_R(t)  = \kappa' \left[c(t) -c_{\rm out} \left(1+\frac{l_{_{\rm GT}}}{R(t)}\right) \right]
\ ,
\eeq
where $\kappa'$ is a  constant of dimension [length]/[time],
 which is proportional to the effective adsorption (desorption) rate of a polymer into (out of) the drop caused by the mismatched boundary conditions at the drop's interface. The rate is effective in the sense that it corresponds to the statistical average of many microscopic adsorption of desorption events that occur at the interface.

In an $N$-drop system, the rate of change of the drops' radii  are thus:
\beq
\frac{\dd {R}_i(t)}{\dd t}= \frac{J_{R_i}(t)}{c_{\rm in}}=\kappa \left(\frac{1}{R_c(t)} - \frac{1}{R_{i}(t)}\right)
\ ,
\label{eq:Ri}
 \eeq
 for $1 \leq i \leq N$,  where $\kappa \equiv \kappa' l_{_{\rm GT}} c_{\rm out}/c_{\rm in}$ and 
 \beq
 \label{eq:Rc}
 R_c(t) = \frac{l_{_{\rm GT}}}{c(t)/c_{\rm out} -1}
 \ ,
 \eeq which is termed the {\it critical radius} since all drops of radii below $R_c$ shrink, and vice versa. The set of  $N$ differential equations  (\ref{eq:Ri})  are coupled through the mass conservation condition in (\ref{eq:c}) \& (\ref{eq:Rc}).

Surprisingly, the model equations of conversion-limited phase separation (\ref{eq:Ri}) can in fact be mapped onto a grain growth model in a single-phase material in three dimensions studied by Hillert \cite{hillert_actamet65}. In this mapping, the parameter  $\kappa$ corresponds to
the proportionality constant that relates the velocity of the advancement of a grain boundary and the local curvature of the grain boundary. Interestingly, discrepancies between actual grain growth and the the Hillert's model is known and is caused by the mean-field assumption when calculating the grain boundary curvature \cite{darvishi_actamat12}. Such a deficiency does not occur in our system since the drops remain spherical at all times.

\begin{figure*}
	\begin{center}
		\includegraphics[scale=.72]{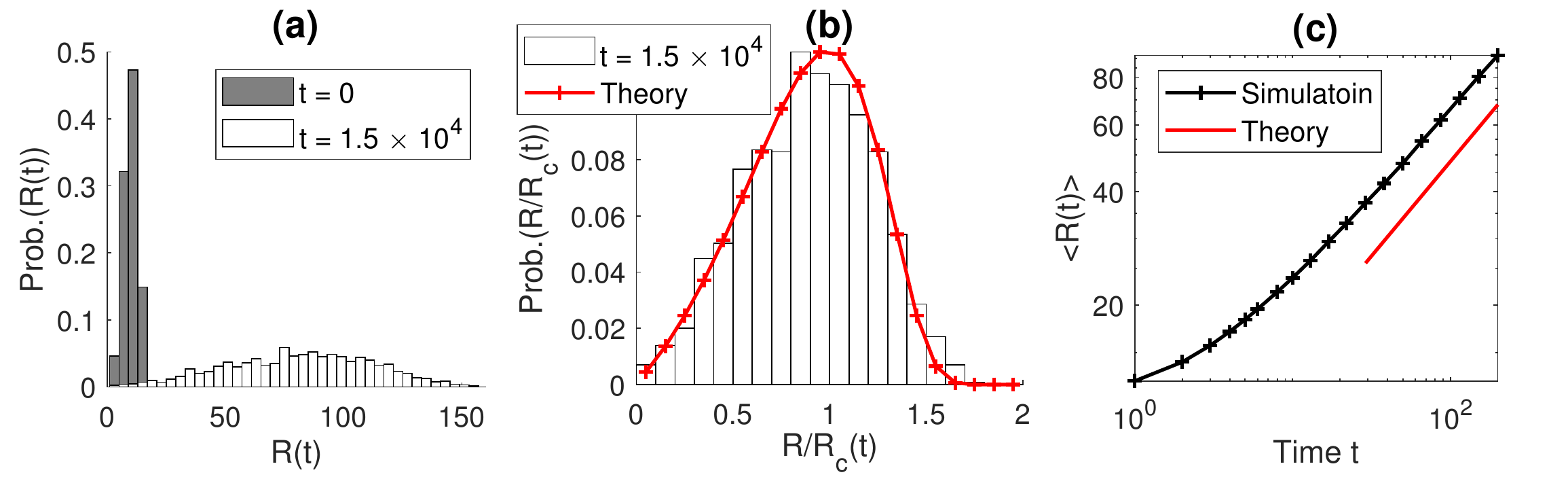}
	\end{center}
	\caption{Simulation vs.~Theory. (a) The broadening of the drop size distribution that starts with a Gaussian distribution  (mean = 10, standard deviation = 3), but truncated so that the random variable is always positive (grey). The other parameters are $N=5\times 10^5$, $\kappa =1$, $c_{\rm out}=1$, $c_{\rm in} =20$, $c_{\rm tot}= c_{\rm in} +99 c_{\rm out}\left[1+\la R(t=0)\ra^{-1}\right]$.  Note that the late-stage universal behavior discussed here are not dependent on the choice of the parameters.
	(b) The drop size distribution (at $t=1.5 \times 10^5$) and (c) the temporal evolution of  $\la R(t) \ra$ show good agreements with the theoretical predictions  (\ref{eq:sizedist}) and (\ref{eq:Rscaling}), respectively.
	}
	\label{fig:results}
\end{figure*}

\section{Conversion-limited vs.~Lifshitz-Slyozov--Wagner}
In the Lifshitz-Slyozov--Wagner (LSW) scheme, the prefactor $\kappa$ in (\ref{eq:Ri}) is replaced by $D l_{_{\rm GT}} c_{\rm out}/[c_{\rm in} R(t)]$ \cite{lifshitz_jpcs61,wagner61, weber_rpp19}, where $D$ is the diffusion coefficient of the molecule in the dilute phase. Comparing these two quantities, one expects that the conversion-limited scheme is valid when the following condition is satisfied:
\beq
\label{eq:condition}
\kappa\ll \frac{D}{\la R(t) \rangle}
\ .
\eeq

 For a typical protein of linear dimension around 1 nm, $D$ is of the order $10^7$ nm$^2$/s \cite{milo_b15}. Taking the typical condensate size to be in the order of 1000 nm, the conversion-limited regime is expected to be valid if $\kappa \ll 10^4$ nm/s. As an example, a recent  study of P granules in the single-cell embryonic stage estimated
that   the parameter $\kappa$ is of the order  1 nm/s  \cite{folkmann_bio21}. Therefore, the conversion-limited scheme is appropriate for that system.

\section{Scaling law \& universal size distribution}
Having  reviewed the physics underlying  the conversion-limited scheme and compared the model equations  to those of the LSW scheme, I will now elucidate analytically the  universal behavior of coarsening in the asymptotic long time limit. While it is unclear whether this asymptotic regime is relevant to intracellular condensates in vivo, 
the emergent universal physics can clearly be tested in controlled experiments, and be applicable to diverse natural, re-constituted, or synthetic phase separating systems,
 such as various phase separating polymers with polyvalent binding sites.

As noted before, the model equations can be mapped onto a grain growth model in a single-phase material in three dimensions \cite{hillert_actamet65}, where the coarsening behavior has also been analyzed analytically. For completeness, I will present a similar derivation of the analytical results in this section, while highlighting  how they differ from the predictions  of the  standard theories of phase separation.

As the identities of individual drops are irrelevant, 
 I will start by  focusing on the following time-dependent drop size distribution function:
\beq 
 n(R,t) = \frac{1}{V}\sum_{i=1}^N \delta (R-R_i(t))
 \ .
 \label{eq:n}
\eeq
 
 Now, recall that in the asymptotic long time limit of the LSW regime, 
the distribution function $n$ approaches the  scaling form \cite{lifshitz_jpcs61,pitaevskii_b81}:
\beq
\lim_{t \rightarrow \infty} n_{\rm LSW}(R,t) = \left[R_c(t)\right]^{-4} g_{\rm LSW} \left( z \right)
\ ,
\eeq
where $z(t)\equiv R/R_c(t)$ and $g_{\rm LSW}$ is a dimensionless scaling function given by
\beq
g_{\rm LSW}(z) \propto \frac{z^2\exp \left(1-\frac{3}{3-2z}\right)}{(1+\frac{z}{3})^{7/3}(1-\frac{2z}{3})^{11/3}}
\ ,
\eeq
for $0\leq z \leq 3/2$. In other words, in the  long time regime, the drop size probability distribution, once renormalized by the critical radius,  is temporally invariant.
I will assume that the same scale invariant structure remains true for the conversion-limited scheme. Specifically,  I will
use the {\it ansatz}:
 \beq
 n(R,t)= \left[R_c(t)\right]^{-4} g(z(t))
 \label{eq:def_g}
 \eeq
 where  $g(.)$ is a dimensionless function.  
In fact, the ansatz (\ref{eq:def_g}) is, as in the LSW scheme, an inevitable outcome of mass conservation in the system. The demonstration of this asymptotic behavior mirrors exactly that of the Lifshitz-Slyozov--Wagner theory \cite{hillert_actamet65,lifshitz_jpcs61,pitaevskii_b81}, and therefore will not be repeated here.
 
I will now calculate $\pp_t n$ by using first the definition of $n$ in (\ref{eq:n}) to get
\begin{subequations}
\label{eq:lhs}
\begin{align}
\pp_t n(R,t) =& -\frac{\pp}{\pp R} \left[\kappa\left(\frac{1}{R_c(t)} -\frac{1}{R} \right) n(R,t)\right] 
\\
=& -\frac{\kappa}{R_c^6} \left[\frac{g}{z^2 } +\left(1 -\frac{1}{z} \right) g' \right]
\ .
\end{align}
\end{subequations}
In the second equality above, I have replaced all $R$ and $n$ in the last expression  by $z$ and $g$ using the ansatz  (\ref{eq:def_g}).

Calculating $\pp_t n$ for a second time, but using  the ansatz 
 (\ref{eq:def_g}) directly instead, we get
 \beq
\pp_t n(R,t) = -\frac{1}{R_c^{5}} \frac{\dd R_c}{\dd t} \left
( 4g  +z g'\right)
\label{eq:rhs}
\ ,
\eeq
where $g' = \dd g /\dd z$. 

Equating (\ref{eq:rhs}) and (\ref{eq:lhs}b), and then separating the $R_c(t)$ and $t$ on one side, and $g(z)$ and $z$ on the other, we have
\beq
\label{eq:mainodes}
\frac{R_c}{\kappa} \frac{\dd R_c}{\dd t}  = A =\frac{g/z^2+(1-1/z) g'}{4g+zg'} 
\ ,
\eeq
where $A$ is a constant to be determined. 

From the first equality, we can see that the critical radius $R_c$ scales like $t^{1/2}$.
Therefore, $R_c$ increases much faster than the LSW scaling law: $R^{\rm LSW}_c(t) \sim t^{1/3}$. 
Since the  scaling law corresponding  to coalescence-driven coarsening is identical to that of the LSW scheme \cite{siggia_pra79,cates_r17}, {\it the increase in the power law in the conversion-limited scheme justifies the  neglect of coalescence-driven coarsening in the late stage}. 

Note  that a generalized LSW scheme that accounts for a concentration-dependent mobility term can also  lead to  modified scaling laws \cite{bray_prb95}. However, the models considered there always lead to a slowing down of coarsening compared to the LSW scheme.

Focusing now on the drop sizes, solving the differential equation  from the second equality in (\ref{eq:mainodes}) leads to the universal normalized distribution:
\beq
\label{eq:sizedist}
g(z) \propto \frac{z\exp \left(-\frac{6}{2-z}\right)}{(2-z)^5}
\ ,
\eeq
for $0\leq z \leq 2$.
By calculating the average of the distribution, we find the following:
\beq
\label{eq:Rscaling}
 \la R(t) \ra= \frac{24}{27}R_c(t) \sim t^{1/2}
\ ,
\eeq
which is again different from the LSW theory: $ \la R(t) \ra=R^{\rm LSW}_c(t)$. 
 Note that the expressions in (\ref{eq:sizedist}, \ref{eq:Rscaling}) are equivalent to those of grain size distribution and growth growth rate in the late stage found in \cite{hillert_actamet65}.
Incidentally, the $t^{1/2}$ scaling law (\ref{eq:Rscaling})  also coincides with the scaling law expected from the coarsening in non-conserved system undergoing phase ordering \cite{bray_advphys02,krapivsky_b10}.

 \fig \ref{fig:results} shows the verifications of  all theoretical predictions  by numerically solving the model equations  (\ref{eq:c}), (\ref{eq:Ri}) \& (\ref{eq:Rc}). The details of the numerical procedure is given in Appendix.

\section{Discussion \& Outlook}
In summary, I have discussed the physics underlying the late-stage coarsening of a phase separating system under the 
   conversion-limited scheme, and elucidated the scaling law and universal drop size distribution in this regime. Besides the change of the coarsening power law, the conversion-limiting scheme is also arguably more {\it universal} compared to
  the  LSW scheme due to the uniform concentration in the dilute phase because (i)  the spatial correlation of drops is irrelevant and thus the ``mean-field" scenario considered here is exact \cite{yao_prb92}, and (ii) the universal behavior is independent of the spatial dimension \cite{sanmiguel_pra85,rogers_prb89}.

Referring back to the condition (\ref{eq:condition}) under which the conversion-limited regime is valid, it is clear that as the average drop size  grows ($\la R(t) \ra \rightarrow \infty$), the system will transition into the LSW scheme eventually. In other words, the asymptotic results described here are strictly speaking only applicable to some intermediate scaling regime. However, depending on the model parameters, this intermediate scaling regime can be extremely long.  For instance, using  the  parameters estimated in an in vivo study of  P granules in the one-cell embryo \cite{folkmann_bio21}, one finds that the transition from the conversion-limited regime to the LSW regime  occurs when $\la R \ra \approx 10^7$ nm, which corresponds to an intermediate scaling regime that spans over 10,000 years.

A final surprise here is that the universal behavior is uncovered in a system that is, while clearly motivated by a biological system, purely thermal. This is in contrast to, e.g., another class of living matter-motivated systems -- polar active matter, in which diverse novel universality classes emerge from the  non-equilibrium nature of the systems \cite{vicsek_prl95,toner_prl95,toner_pre98,chen_njp15,toner_prl18,toner_pre18,
chen_prl20,chen_pre20}. Overall, this work highlights once again that biological systems constitute a fertile ground for  novel physics \cite{lee_jpd19}.

In terms of outlook, 
one potentially interesting direction will be to consider the impact of having multiple co-existence phases  \cite{sear_prl03,jacobs_biophysj17,mao_softmatt19,jacobs_prl21} on the coarsening behavior. Although for the standard LSW theory,  having multiple phases does not seem to influence the asymptotic scaling behavior \cite{jeppesen_prb93,das_pre02}, it would be interesting to see whether the growth of drops of distinct phases
in the intermediate stage will be affected by potential competition over shared components.

Another interesting direction  will be the study of how the  universal behavior of phase separation  can be impacted by non-equilibrium processes in cells, which can include the active motility of the constituent components (e.g., due to molecular motors) and driven chemical reactions (e.g., ATP-driven enzymatic reactions).

\appendix*
\section{Numerical procedure}
\label{app}
To study the temporal evolution of the drop sizes (Fig.~2), the following numerical procedure is used:
\begin{enumerate}
\item
{\it Initiation.}
 At time  $t=0$,  Gaussian  variables (with mean = 10, standard deviation = 3)  are repeatedly drawn until $N$ positive numbers are obtained. These will be the initial radii of the $N$ drops $R_i(t=0), i=1,\ldots, N$.
 \item
{\it Evolution.} The drop sizes are updated according to the followings:
\beq
R_i(t+\tri t ) =R_i(t) + \tri t \times \kappa \left(\frac{1}{R_c(t)} - \frac{1}{R_{i}(t)}\right)
\ ,
\eeq
where $\tri t = 0.01$, and $R_c(t)$ is calculated using  (\ref{eq:Rc}). 
Any $R_i(t)$ that drops below $10^{-3}$ will be taken out of the system.
Time $t$ is then updated to $t+\tri t$.
\end{enumerate}

The  model parameters used to simulate the results presented in \fig.~2 are shown in the caption. However, the universal size distribution (\ref{eq:sizedist}) and scaling law (\ref{eq:Rscaling}) are not dependent on the choice of the parameters in the late stage.

\begin{acknowledgments}
CFL thanks Andrew Folkmann, Andrea Putnam, and Geraldine Seydoux for many stimulating discussions.
\end{acknowledgments}


%

\end{document}